\documentstyle[preprint,aps,epsfig]{revtex}
\tighten
\def\bea{\begin{eqnarray}}
\def\beann{\begin{eqnarray*}}
\def\beq{\begin{equation}}
\def\eea{\end{eqnarray}}
\def\eeann{\end{eqnarray*}}
\def\eeq{\end{equation}}
\def\nn{\nonumber}

\newcommand{\bcdot}{\bbox{\cdot}}

\newcommand{\btimes}{\bbox{\times}}

\newcommand{\bkhat}{\bbox{\hat k}}

\newcommand{\bsigma}{\bbox{\sigma}}

\newcommand{\bepsilon}{\bbox{\epsilon}}

\begin{document}
\headheight1.2cm
\headsep1.2cm
\baselineskip=20pt plus 1pt minus 1pt

%
\preprint{MKPH-T-97-30}
\title{Dispersion relations and the spin polarizabilities of the nucleon}
\author{D. Drechsel, G. Krein\thanks{\noindent
Alexander von Humboldt Research Fellow\hfil\break
Permanent Address: Instituto de F\'{\i}sica Te\'{o}rica, Universidade Estadual
Paulista\hfil\break
Rua Pamplona, 145 - 01405-900 S\~{a}o Paulo, SP - Brazil}, 
and O. Hanstein\thanks{\noindent Supported by Deutsche Forschungsgemeinschaft
(SBF 201)}  }
\address{Institut f\"{u}r Kernphysik, Universit\"{a}t Mainz, 55099 
Mainz, Germany }
\maketitle
\begin{abstract}
A forward dispersion calculation is implemented for the spin 
polarizabilities $\gamma_1, \cdots ,\gamma_4$ of the proton and the neutron. 
These polarizabilities are related to the spin structure of the nucleon at low 
energies and are structure-constants of the Compton scattering
amplitude at ${\cal O}(\omega^3)$. In the absence of a direct experimental
measurement of these quantities, a dispersion calculation serves the purpose
of constraining the model building, and of comparing with recent calculations 
in heavy baryon chiral perturbation theory. 
\end{abstract}
\vspace{5.50cm}
\noindent{PACS NUMBERS: 13.60.Fz,11.55.Fv,14.20.Dh,13.60.Le,13.88.+e }

\vspace{0.3cm}
\noindent{KEYWORDS: Compton scattering, dispersion relations, nucleon 
polarizabilities, spin content of the nucleon}
\newpage
\hspace*{-\parindent}{\bf 1. Introduction}.\hspace*{\parindent}

The response of the internal degrees of freedom of the nucleon to an external
electromagnetic field can be parametrized in terms of the structure 
dependent polarizabilities. The classic process for studying such quantities
is Compton scattering at energies below the resonance region. In an expansion 
of the Compton scattering amplitude in powers of the incident photon energy 
$\omega$, the ${\cal O}(1)$ and ${\cal O}(\omega)$ terms depend only on the 
mass $M$, the electric charge $e$ and the anomalous magnetic moment $\kappa$ 
of the nucleon. Therefore, no information on the excitation of the internal 
degrees of freedom can be obtained up to ${\cal O}(\omega)$; it is only at 
${\cal O}(\omega^2)$ that the amplitude becomes sensitive to the internal 
excitation of the nucleon. At ${\cal O}(\omega^2)$, the amplitude is 
parametrized in terms of the electric ($\alpha$) and magnetic ($\beta$) 
polarizabilities~\cite{klein}, which describe the deformation of the system 
in the presence of a static electric and magnetic field. At 
${\cal O}(\omega^3)$, as demonstrated recently by Ragusa~\cite{ragusa}, four 
new polarizabilities appear, $\gamma_1, \gamma_2, \gamma_3$, and $\gamma_4$, 
the ``spin polarizabilities". As explained by Ragusa, these can be 
interpreted as the response to the external fields of a magnet that 
has an internal structure. For related previous work on the 
${\cal O}(\omega^3)$ spin structure of the Compton scattering, see 
Ref.~\cite{previous}.

Contrary to $\alpha$ and $\beta$,  none of the $\gamma_i$'s has been measured 
experimentally, although there has been a proposal for that 
purpose~\cite{miskimen}. Only the spin-dependent polarizability for
forward scattering involving three of the spin polarizabilities, 
$\gamma=\gamma_1-\gamma_2-2\gamma_4$,  has been experimentally constrained 
through an analysis involving photoproduction multipoles~\cite{swk}, 
with the result $\gamma=-1.3 $ (here and thereafter we quote all 
polarizabilities in units of $10^{-4}\,$fm$^4$). An earlier multipole 
analysis~\cite{toyla} obtained $\gamma =-1.0$ .

On the theoretical side, there are recent calculations of these 
quantities using chiral perturbation theory (ChPT). In Ref.~\cite{BKKM}, 
$\gamma$ was evaluated in a ${\cal O}(p^3)$ calculation in heavy baryon ChPT 
(HBChPT), with the result:
\beq
\gamma= \frac{e^2}{4\pi} \frac{g^2_A}{24\pi^2 F^2_{\pi} m^2_\pi} =
4.6 ,
\label{gam1st}
\eeq
where $F_{\pi}=92.4$ MeV, $m_{\pi}=137$~MeV, and 
$g_A = 1.26$. This prediction is in clear
disagreement with the multipole analysis. Sub-leading effects evaluated 
within a relativistic ChPT calculation at the one-loop level diminish the
discrepancy, yielding~\cite{BKKM} $\gamma^{1-loop} = 2.2 $. Moreover,
inclusion of the $\Delta(1232)$ resonance through higher order 
contact terms within the same approach~\cite{BKKM} revealed large 
effects on $\gamma$, with opposite sign to the one-loop result, 
$\gamma^{\Delta}=-3.7$. The net result in the relativistic calculation,
\beq
\gamma= + 2.2\,(\text{1-loop}) - 3. 7 \, (\Delta) = -1.5,
\label{numbers0}
\eeq
is very close to the one found with the multipole analysis mentioned above.  

The individual spin polarizabilities were calculated in Ref.~\cite{BKMrev} 
within the same HBChPT approach as in Ref.~\cite{BKKM}. An interesting result 
of this calculation is that the values of $\gamma_1, \gamma_3$ and $\gamma_4$ 
are completely dominated by the contribution of the $t$-channel 
$\pi^0$-exchange (Wess-Zumino-Witten term), while this contribution cancels 
in the expression for $\gamma$. Specifically, the expressions obtained for the
$\gamma_i$'s are
\bea
&&\gamma_1 = \frac{e^2}{4\pi} \frac{g_A}{4\pi^2 F^2_{\pi} m^2_\pi}
\left(-1+\frac{g_A}{6}\right),\hspace{1.75cm}
\gamma_2 = \frac{e^2}{4\pi} \frac{g_A}{4\pi^2 F^2_{\pi} m^2_\pi}
\left(0+\frac{g_A}{12}\right),\nn\\
&&\gamma_3 = \frac{e^2}{4\pi} \frac{g_A}{4\pi^2 F^2_{\pi} m^2_\pi}
\left(\frac{1}{2}+\frac{g_A}{24}\right),\hspace{2.0cm}
\gamma_4 = -\gamma_3.
\eea

Very recently, in Ref.~\cite{HHK} the $\Delta(1232)$ was introduced as an 
explicit degree of freedom in HBChPT. In this approach a new dimensionful 
parameter enters, $\Delta = M_{\Delta} - M$, and a consistent HBChPT 
formalism can be set up according to an ${\cal O}(\epsilon^n)$ power counting 
scheme, where 
$\epsilon$ denotes a small scale (a soft momentum, 
$m_{\pi}$ or $\Delta$). In this formalism the previous ${\cal O}(p^n)$ HBChPT 
results~\cite{BKMrev} are exactly reproduced, and the terms that come from 
treating the $\Delta(1232)$ as an explicit degree of freedom can be clearly 
identified. The value for $\gamma$ obtained to ${\cal O}(\epsilon^3)$ is 
\bea
\gamma &=& \frac{e^2}{4\pi} \left\{\frac{g_A^2}{24\pi^3 F^2_{\pi} m^2_\pi} 
- \frac{4b^2_1}{9M^2} \frac{1}{\Delta^2}
- \frac{g^2_{\pi N\Delta}}{54\pi^2 F^2_{\pi}}
\left[ \frac{\Delta^2 + 2m^2_{\pi}}{(\Delta^2 - m^2_{\pi})^2}
-\frac{3m^2_{\pi}\Delta}{(\Delta^2 - m^2_{\pi})^{5/2}}\ln R\right]\right\},
\label{gammaHHK}
\eea
where $R=\Delta/m_{\pi}+[(\Delta/m_{\pi})^2-1]^{1/2}$, $g_{\pi N\Delta}$ is
the pion-nucleon-$\Delta(1232)$ coupling constant, and $b_1$ is a finite
${\cal O}(\epsilon^2)$ counterterm. The terms in Eq.~(\ref{gammaHHK}) 
correspond, in order, to nucleon-loop (N-loop), delta-pole ($\Delta$-pole)
and delta-loop ($\Delta$-loop). As remarked in Ref.~\cite{HHK}, both 
$g_{\pi N\Delta}$ and $b_1$ are not well constrained by the experiment.
However, in using the values $g_{\pi N\Delta}=1.5\pm 0.2$
and $b_1 = -(2.5 \pm 0.35)$~\cite{DMW} in Eq.~(\ref{gammaHHK}), Hemmert
et al.~\cite{HHK} obtained
\beq
\gamma = 4.6 \,(\text{N-loop}) - 4.0 \,(\Delta-\text{pole}) - 0.4 
\,(\Delta-\text{loop})=+0.2.
\label{numbers1}
\eeq
As in the calculation of Ref.~\cite{BKKM}, the contribution of the 
$\Delta(1232)$ is large and negative, but cannot account for the value 
$\gamma = -1.3$ found from the multipole analysis. Unfortunately, also the 
values for the electric and magnetic polarizabilities, $\alpha$ and $\beta$, 
are larger than the experimental data.

More recently, Holstein et al.~\cite{argue}~\cite{HHKK} have argued that the 
values of $g_{\pi N\Delta}$ and $b_1$ to be used should be determined, for 
consistency, within the same ``small scale expansion" of the HBChPT 
calculation, and not within a relativistic Born model. In such a case, 
the values for these quantities are approximately 50\% smaller than the ones 
used above. The consequence of this is that $\alpha$ and $\beta$ decrease 
significantly, but on the other hand, the value for $\gamma$ increases
significantly:
\beq
\gamma = 4.6 \,(\text{N-loop}) - 2.4 \,(\Delta-\text{pole}) - 0.2 
\,(\Delta-\text{loop})=+2.0 ,
\label{numbers2}
\eeq
in contradiction with the multipole analysis of Ref.~\cite{swk}. 
The effect of the $\Delta$ pole is also quite large for some of the
$\gamma_i$'s. In Table~I we show the results for the $\gamma_i$'s using for 
$g_{\pi N\Delta}$ and $b_1$ the values suggested by Holstein et 
al.~\cite{argue}~\cite{HHKK}.

As mentioned above, there are no direct measurements of the spin 
polarizabilities. Nevertheless, as in the case of 
$\gamma$~\cite{swk}~\cite{toyla}, 
an estimate for the $\gamma_i$'s can be obtained using the experimentally 
determined amplitudes for pion photoproduction in a dispersion integral. 
However, there are additional contributions from $t$-channel processes that
are not well constrained experimentally, corresponding to possibly large 
high-energy contributions to the dispersion integrals, similar to the effects 
of the Wess-Zumino-Witten term in the case of ChPT. In this sense, only the 
low energy part of the Compton process, i.e. contributions of the pion cloud 
and the low-lying resonances, can be at present constrained by a dispersion 
calculation. 

In this Letter we present the results of a dispersion calculation for the
spin polarizabilities. In order to minimize uncertainties with respect to
the high-energy behavior of the Compton amplitude, we follow the approach of 
L'vov et al.~\cite{lvov}. In this approach one uses a finite-size contour in 
the complex plane rather than improving the convergence of the integral by 
subtractions. The contributions arising from the contour are expressed in 
terms of low-mass meson exchanges in the $t$-channel, as will be discussed
in the next section. We use pion photoproduction multipoles from a recent 
analysis based on fixed~$t$ dispersion relations and unitarity by Hanstein, 
Drechsel and Tiator (HDT)~\cite{HDT}. This analysis is limited to a
maximum photon laboratory energy of the order of 500 MeV. In order to
investigate the sensitivity of our results to the photoproduction
amplitudes at higher energies, we also employ the multipole analysis
VPI-SP97K of the Scattering Analysis Interactive Dial-in program,
SAID~\cite{SAID}. In Section~3 we present our numerical results and
compare them with the results of ChPT.  Some conclusions and
perspectives are presented in Section~4.

\newpage \hspace*{-\parindent}{\bf 2. Dispersion relations for the
spin polarizabilities}.\hspace*{\parindent}

In the c.m. system the $T$-matrix for Compton scattering on the
nucleon can be written, in the Coulomb gauge, in terms of six
amplitudes that are functions of the energy of the incident photon,
$\omega$, and the scattering angle, $\theta$, as~\cite{Pr}: 
\bea 
&&T = \chi^{\dag}_{s'}\left\{
\bepsilon'^*\bcdot\bepsilon\, \bar A_1(\omega,\theta) +
\bepsilon'^*\bcdot\bkhat\, \bepsilon\bcdot\bkhat'\,\bar A_2(\omega,\theta) + 
i\bsigma\bcdot(\bepsilon'^*\btimes\bepsilon)\,\bar A_3(\omega,\theta) +
i \bsigma\bcdot(\bkhat'\btimes\bkhat)\bepsilon'^*\bcdot\bepsilon\,
\bar A_4(\omega,\theta)\right.\nn\\ 
&&+ \left.i \bsigma\bcdot[(\bepsilon'^*\btimes\bkhat)\bepsilon\bcdot\bkhat'
-(\bepsilon\btimes\bkhat')\bepsilon'^*\bcdot\bkhat]\,\bar
A_5(\omega,\theta) + i
\bsigma\bcdot[(\bepsilon'^*\btimes\bkhat')\bepsilon\bcdot\bkhat'
-(\bepsilon\btimes\bkhat)\bepsilon'^*\bcdot\bkhat]\,\bar
A_6(\omega,\theta) \right\} \chi_{s}, 
\label{TAbar} 
\eea 
where $\bkhat,\bepsilon\;(\bkhat',\bepsilon')$ are the direction and the
polarization vector of the incident (final) photon, $\chi_s$ ($\chi_{s'}$)
is the initial (final) nucleon spinor, and  $\bsigma$ is
the Pauli spin matrix. For convenience, we write the amplitudes $\bar
A_i$ as a sum of the Born and non-Born contributions, $\bar A_i = \bar
A^{\text{Born}}_i + \bar A^{\text{nB}}_i$. Ragusa's spin
polarizabilities can be expressed in terms of the non-Born amplitudes
$\bar A^{\text{nB}}_i$'s by~\cite{BKMrev} 
\bea 
&&\gamma_1 =\frac{1}{4\pi} \frac{1}{6} \frac{\partial^3}{\partial \omega^3}
\left[\bar A^{\text{nB}}_3(\omega,0) + \bar A^{\text{nB}}_4 (\omega,0) + 2
\bar A^{\text{nB}}_5(\omega,0)\right]_{\omega=0},\hspace{0.5cm}
\gamma_2 = \frac{1}{4\pi} \frac{1}{6} \frac{\partial^3}{\partial
\omega^3} \left[\bar A^{\text{nB}}_4
(\omega,0)\right]_{\omega=0},\nn\\ 
&&\gamma_3 = \frac{1}{4\pi}
\frac{1}{6} \frac{\partial^3}{\partial \omega^3} \left[\bar
A^{\text{nB}}_6 (\omega,0)\right]_{\omega=0},\hspace{2.5cm} \gamma_4 =
\frac{1}{4\pi} \frac{1}{6} \frac{\partial^3}{\partial \omega^3}
\left[\bar A^{\text{nB}}_5 (\omega,0)\right]_{\omega=0}.
\label{gamAbar} 
\eea 

On the other hand, the scattering amplitude can be written in terms
of the Hearn-Leader (HL)~\cite{HL} Lorentz-scalar amplitudes
$T_i(\nu,t), i=1 \cdots 6$, as 
\beq T= \epsilon^{\mu} \, \bar u_{s'}(p')\left[\,
\sum_{i=1}^6 I^{i}_{\mu\nu} T_i(\nu,t) \,\right] u_s(p)\,
\epsilon^{\nu}, \label{TLorScal} 
\eeq 
where $\epsilon^{\mu}$ and $\epsilon^{\nu}$ are the polarization four-vectors 
of the nucleon, $\bar u_s(p)$ and $u_s(p)$ are Dirac spinors, 
and $I^{i}_{\mu\nu}$ are tensors that depend on Dirac $\gamma$ matrices and 
the initial and
final momenta of the photon and the proton. From arguments based on the
asymptotic behavior of Regge trajectories, some of the HL amplitudes
appear to have a bad convergence for fixed $t$ and large $\nu$. The
traditional approach in dispersion theory is to implement a
subtraction at threshold for these amplitudes with a bad asymptotic
behavior, at the cost of introducing subtraction functions that in
most cases are unknown. However, there is an additional problem with
the HL amplitudes $T_i(\nu,t)$, in that they have to fulfil kinematic
constraints because the tensors $I^i_{\mu\nu}$ develop singularities
at forward and backward scattering angles~\cite{BT}. It is possible to
avoid kinematic constraints and singularities by using appropriate
combinations of the amplitudes, as for example the ones of Bardeen and
Tung~\cite{BT}. One natural way to proceed is the one followed by Pfeil,
Rollnik and Stankowski~\cite{bonn}, who used a partial wave
decomposition of the HL amplitudes. While the constraints can be
automatically satisfied within this procedure, it has the disadvantage
of introducing a possible violation of the $s-u$ crossing
symmetry. Another approach has been introduced by L'vov et. al~\cite{lvov},
who expresses the dispersion contribution as the sum of a finite-range
integral from threshold $\nu_{\text{thr}}$ up to a maximum value
$\nu_{\text{max}}$, and the contributions from higher energies in
terms of $t$-channel poles. Thus, the real parts of the amplitudes are
written as a sum of the real pole Born terms, the contribution of the
finite-range dispersion integral, and an asymptotic $t$-channel
contribution (we use the conventions and notations of
Ref.~\cite{lvov}): 
\beq \text{Re} A_i(\nu,t) =
A^{\text{Born}}_i(\nu,t) + \frac{2}{\pi}\text{P}
\int^{\nu_{\text{max}}}_{\nu_{\text{thr}}}\; \text{Im}A_i(\nu',t)\;
\frac{\nu'\,d\nu'}{\nu'^{2}-\nu^2} + A^{\text{as}}_i(t).  
\label{ReA}
\eeq 
The amplitudes $A_i(\nu,t)$ are appropriate combinations of the
HL amplitudes, and are free of kinematic constraints and
singularities. Two of these six amplitudes, $A_1$ and $A_2$, appear to
have a bad convergence behavior for high $\nu$, at fixed $t$. The
$t$-channel contribution to $A^{\text{as}}_1(t)$ and
$A^{\text{as}}_2(t)$ are modeled by $\sigma$ and $\pi^0$ exchanges,
respectively, and for the remaining amplitudes the $t$-channel
contributions seem to give negligible contribution at low energies.

In order to evaluate the spin polarizabilities in terms of the
dispersion integral, one needs the relation of the amplitudes $\bar
A_i(\omega,\theta)$ of Eq.~(\ref{TAbar}) to the Lorentz-scalar
amplitudes $A_i(\nu,t)$. The relation is easily obtained comparing
Eq.~(\ref{TAbar}) with the low energy expansion of the Compton
scattering amplitude written in terms of the non-Born contributions to
the Lorentz-scalars $A_i(\nu,t)$, with the result~\cite{bab}~\cite{us} 
\bea
&&\gamma_1 = - \frac{1}{4\pi} \frac{1}{M}\left[ A^{\text{nB}}_2(0,0) -
A^{\text{nB}}_4(0,0) + A^{\text{nB}}_5(0,0)\right],\nn\\ 
&&\gamma_2 = + \frac{1}{4\pi} \frac{1}{M}\left[ A^{\text{nB}}_5 (0,0) -
A^{\text{nB}}_6 (0,0)\right],\nn\\ 
&&\gamma_3 = + \frac{1}{8\pi}
\frac{1}{M}\left[ A^{\text{nB}}_2 (0,0) - A^{\text{nB}}_4 (0,0) -
A^{\text{nB}}_6 (0,0)\right],\nn\\ &&\gamma_4 = - \frac{1}{8\pi}
\frac{1}{M}\left[ A^{\text{nB}}_2(0,0) + A^{\text{nB}}_4 (0,0) + 2
A^{\text{nB}}_5 (0,0)- A^{\text{nB}}_6 (0,0) \right] .
\label{gamALor} 
\eea 
From these expressions one sees that the
polarizabilities $\gamma_1$, $\gamma_3$ and $\gamma_4$ depend on the
amplitude $A_2$.  As we discussed previously, this amplitude has a bad
asymptotic behavior for fixed $t$ and large $\nu$ and not
well-constrained $t$-channel contributions. One way to avoid such
uncertainties is to consider appropriate combinations of the
$\gamma_i$'s to which $A_2$ does not contribute. In addition to the
combination that leads to $\gamma$, there are more of such
combinations, which will be discussed in the next section.

\newpage \hspace*{-\parindent}{\bf 3. Numerical results}.\hspace*{\parindent}

In the dispersion integral Eq.~(\ref{ReA}) we saturate the imaginary
parts by one-pion photoproduction amplitudes. We neglect intermediate
states with more pions and heavier mesons. The contributions of such
states to $\gamma$ have been estimated in Ref.~\cite{swk} to be relatively
small. As stated above, we employ two sets of photoproduction
amplitudes, the ones obtained with a fixed $t$ dispersion relations
(HDT) and the ones from the SAID program.  Our results using the
dispersion relations (DR) for the spin polarizabilities of the proton
and the neutron are presented in Tables~I,~II,~and~III.

The results in Table I are obtained by evaluating the integral in
Eq.~(\ref{ReA}) with the HDT~\cite{HDT} multipoles up to a
$\nu_{\text{max}}=E^{\text{max}}_{\gamma}+t/4M$, with $t=0$ and
$E^{\text{max}}_{\gamma}=500$ MeV. Tables~II~and~III also show
the results for the integration with the VPI-SP97K SAID multipoles up to
$E^{\text{max}}_{\gamma}=$~500 and 1500~MeV, which are denoted by
$\text{SAID}_1$ and $\text{SAID}_2$, respectively. For the $t$-channel
contribution $A^{\text{as}}_2(t)$ we use the parametrization
given~in~Ref.~\cite{lvov}.

In Table~I we show the separate contributions from $\pi^0$ exchange
and the dispersion integrals (``excitation") to the polarizabilities. The
contribution of the $\pi^0$ exchange to Eq.~(\ref{ReA}) is practically
identical to the Wess-Zumino-Witten term of the HBChPT
calculation. Because of the huge contribution from that term, our
global result ("sum") is quite similar to the prediction of
HBChPT~\cite{HHKK}. However, the contributions beyond the anomaly
show significant differences. In Table~II we present
the results for these excitation contributions to the spin
polarizabilities using the HDT and SAID multipoles. One notices that
the HDT and $\text{SAID}_1$ results are fairly similar for all the
$\gamma_i$'s. On the other hand, extending the upper limit of the
integration of the SAID multipoles from $500$~MeV to $1500$~MeV
changes the results for $\gamma^{(p)}_1$, $\gamma^{(p)}_2$, and 
$\gamma^{(p)}_4$ by a factor of the order of 25\%, and changes the sign and 
magnitude of $\gamma^{(p)}_3$. The values of the $\gamma^{(n)}_i$'s are more
stable with respect to the change of the upper limit of the itegration.
We come back to this point shortly ahead.  As already seen
in Table~I, the results from DR are at variance with the predictions
of HBChPT in most cases.

As we discussed above, there exist combinations of the $\gamma_i$'s
that do not depend on the amplitude $A_2$ and thus are not affected by
the badly known high-energy contributions. One of these combinations
is $\gamma=\gamma_1-\gamma_2-2\gamma_4$, and three other combinations
are: 
\beq 
\gamma_{13} = \gamma_1 + 2 \gamma_3, \hspace{1.0cm}
\gamma_{14} = \gamma_1 - 2 \gamma_4, \hspace{1.0cm} \gamma_{34} =
\gamma_3 + \gamma_4.  
\label{combo} 
\eeq 
Note that only two of these combinations are independent, since
$\gamma_{13}-\gamma_{14}=2\gamma_{34}$. The results for $\gamma$, 
$\gamma_{13}$, and $\gamma_{14}$ are presented in Table~III.

Our results for $\gamma^{(p)}$ and $\gamma^{(n)}$ using the SAID
multipoles up to $1500$~MeV are practically identical to the values
obtained for these quantities in Ref.~\cite{swk} using the VPI-FA93
analysis (see Ref.~\cite{swk} for details). We note that the
values of $\gamma^{(p)}$ and $\gamma^{(n)}$ using the VPI-SP97K SAID multipoles
up to $500$~MeV and $1500$~MeV differ by less than 10\%, i.e. 
there are no large contributions to the dispersion integrals from
energies between $500$~MeV and $1500$~MeV. This fact can be understood
as due to the damping factor $1/\omega^3$ in the integrand of the
expression for $\gamma$,
\beq \gamma
=\int^{\infty}_{\omega_{\text{thr}}}
\frac{\sigma_{1/2}-\sigma_{3/2}}{4\pi^2\omega^3}\,d\omega ,
\label{gamma} 
\eeq 
where $\sigma_{1/2}$ and $\sigma_{3/2}$ are the
total photoabsorption cross sections measured with the photon and
nucleon polarizations parallel and antiparallel, respectively. On the
other hand, we find that the value of $\gamma$ calculated with the HDT
multipoles is a factor of two smaller than the value calculated with
the SAID multipoles. This can be traced to differences between the
$E_{0+}$ amplitudes in both analyses. In Figs.~1 (a) and (b) we plot
the contributions of the $E_{0+}$ and $M_{1+}$ multipoles (which are
by far the largest contributors) to the integrand in
Eq.~(\ref{gamma}), and in Fig.~1(c) we plot the total integrand. From
these figures it becomes obvious that the difference comes from the
behavior of the $E_{0+}$ multipoles close to threshold. As remarked in
Ref.~\cite{SAID2}, the SAID multipoles are not meant to be used in the
$\pi^+n$ threshold region. For example, the amplitude $E_{0+}(n\pi^+)$
at threshold is $ 24.9 \times 10^{-3}/m_{\pi^+}$ for SAID and $28.3
\times 10^{-3}/m_{\pi^+}$ for HDT, the latter being much closer to the
threshold value predicted by ChPT~\cite{BKM}, $ 28.4 \times
10^{-3}/m_{\pi^+}$. The differences between the $E_{0+}$ amplitudes of
the HDT and SAID multipoles is of the order of 40\% at the peak value of 
the integrand of $\gamma$. As seen in Fig.~1, this large difference between 
the $E_{0+}$ amplitudes, and a small difference between the $M_{1+}$ 
amplitudes (the SAID value is a little larger than the HDT value), give a net 
effect of 50\% in the final value for $\gamma$.

An interesting comparison can be made between the DR and HBChPT
results by considering in separate the leading contributions to
$\gamma$, which come from the $E_{0+}$, $E_{1+}$, and $M_{1+}$
multipoles. Using the HDT multipoles, we obtain: 
\beq \gamma = + 2.5 \,(E_{0+}) - 3.0 \,(E_{1+},M_{1+}) - 0.1 \,(\text{rest}) 
= - 0.6 ,
\label{numbers3} 
\eeq 
where the last term includes all other partial waves. This shows the strong
cancelation between s-wave pion loop and $\Delta$ resonance
contributions, which is in close correspondence with a similarly
strong cancelation in the ChPT calculations,
Eqs.~(\ref{numbers0}), (\ref{numbers1}), and (\ref{numbers2}). It is amusing
to note that the first two numbers in Eq.~(\ref{numbers3}) are not very
different from the ones in Eq.~(\ref{numbers0}) obtained in the
relativistic ChPT calculation of Ref.~\cite{BKKM}, namely
$\gamma^{1-loop} = 2.2$ and $\gamma^{\Delta}=-3.7$.

To conclude our discussion on the numerical results, we note that the
results in Table III clearly show that when the troublesome amplitude
$A_2(\nu,t)$ is eliminated by taking appropriate combinations of the
$\gamma_i$'s, the results become very stable with respect to the upper
limit of the integral. In fact, similar to the case with $\gamma$, the
integrals for the other combinations of $\gamma_i$'s are almost
completely saturated with an upper limit of $500$~MeV. The large
changes in the $\text{SAID}_1$ and $\text{SAID}_2$ predictions
observed in Table~II can therefore be traced to the presence of the
function $A_2(\nu,t)$ in the expressions of the $\gamma_i$'s. In fact,
inspection of the integrands for the $\gamma_i$'s reveals~\cite{us}
that these still receive sizable contributions from high energies due
to the function $A_2(\nu,t)$.

\vspace{0.3cm} \hspace*{-\parindent}{\bf 4. Conclusions and
Perspectives}.  \hspace*{\parindent}

In this paper we use dispersion relations to estimate the values of
the spin polarizabilities $\gamma_i$ of the nucleon and compare our
results with HBChPT calculations. By taking appropriate combinations
of the $\gamma_i$'s, we are able to minimize uncertainties related to
$t$-channel contributions. The present dispersion calculation predicts, 
in accord with previous calculations~\cite{swk}\cite{toyla}, an opposite 
sign for $\gamma=\gamma_1-\gamma_2-2\gamma_4$ as compared to the HBChPT
calculation.  The same discrepancy is seen for
$\gamma_{14}=\gamma_1-2\gamma_4$. Note that in the HBChPT
calculation~\cite{HHKK}, $\gamma_1$ receives sizable contributions
from $\pi - N$ loops but no $\Delta$-pole contributions, whereas in
$\gamma_2$ and $\gamma_4$ there are cancelations between contributions
from $\pi - N$ loops and the $\Delta$-pole graphs~\cite{HHKK}.  Therefore
the required change of sign in both $\gamma$ and $\gamma_{14}$ seems
nontrivial and highly constrained. In this sense, it would
be extremely interesting to see the outcome of a systematic HBChPT
calculation of these quantities to ${\cal O}(\epsilon^4)$.

Another interesting conclusion of our calculation is that the spin
polarizabilities are very sensitive to the behavior of the
photoproduction multipole $E_{0+}$ near threshold. There is a large
cancelation between the contributions from the $E_{0+}$ and $M_{1+}$
multipoles, and a very precise determination of both of these is
necessary to constrain the values of the spin polarizabilities.

We conclude by remarking that similar to the electric and magnetic
polarizabilities $\alpha$ and $\beta$, the spin polarizabilities
$\gamma_1 \cdots\gamma_4$ contain vital information on the low energy
structure of the nucleon. The experimental determination of these new
polarizabilities is therefore of great interest.

\newpage
\vspace{0.3cm} \hspace*{-\parindent}{\bf Acknowledgments}.\hspace*{\parindent} 

The authors gratefully acknowledge useful discussions with J.~Ahrens, 
A.I.~L'vov, A.~Metz, S.~Scherer, J.~Seaborn, and L. Tiator. The work of GK was 
supported in part by the Alexander von Humboldt Foundation (Germany) and 
FAPESP (Brazil).

%
%
\begin{table}
\caption{Separate contributions to the spin polarizabilities from the HBChPT
calculation of Ref. [12] and the result from the dispersion
calculation using the HDT [14] multipoles (all results are in units of  
$10^{-4}\,$ fm$^4$). }
\begin{center}
\begin{tabular}{c|dcd|ccd}
    {}     &    { }  & HBChPT   &{} & {} & DR (HDT)  & {} \\ 
\tableline
$\gamma_i$ & WZW      & excitation  & sum & $\pi^0$-exchange &
excitation & sum   \\
\tableline
$\gamma^{(p)}_1$ & $-$22.0 & +4.4  & $-$17.6 & $-$22.5  & +5.1     & $-$17.4 \\
$\gamma^{(p)}_2$ &  0      &$-$0.3 & $-$0.3  & 0        & $-$1.1   & $-$1.1  \\
$\gamma^{(p)}_3$ & +11.0   & +1.1  & +12.1   & +11.2    & $-$0.6   & +10.6   \\
$\gamma^{(p)}_4$ & $-$11.0 & +1.3  & $-$9.7  & $-$11.2  & +3.4     & $-$7.9  \\
\tableline
$\gamma^{(n)}_1$ & +22.0   & +4.4  & +26.4  & +22.5  & +6.1   & +28.6  \\
$\gamma^{(n)}_2$ &  0      &$-$0.3 & $-$0.3 &   0    &$-$0.8  &$-$0.8  \\
$\gamma^{(n)}_3$ &$-$11.0  & +1.1  & $-$9.9 &$-$11.2 &$-$0.6  &$-$11.8 \\
$\gamma^{(n)}_4$ & +11.0   & +1.3  & +12.3  & +11.2  & +3.4   & +14.6  \\
\end{tabular}
\end{center}
\end{table}

\vspace{3.5cm}
%
%
\begin{table}
\caption{Excitation contribution to the spin polarizabilities from the 
HBChPT calculation of Ref. [10] and the result from the dispersion
calculation using the HDT, $\text{SAID}_1$, and $\text{SAID}_2$ 
multipoles (all results are in units of  $10^{-4}\,$ fm$^4$). }
\begin{center}
\begin{tabular}{c|cccd}
%
$\gamma_{i}-\text{excit.}$ & HBChPT & DR (HDT) & DR ($\text{SAID}_1$) 
& DR ($\text{SAID}_{2}$)
\\
\tableline
$\gamma^{(p)}_1$ & +4.4  & +5.1   & +4.3  & +3.5   \\
$\gamma^{(p)}_2$ &$-$0.3 &$-$1.1  &$-$1.2 &$-$1.0   \\
$\gamma^{(p)}_3$ & +1.1  &$-$0.6  &$-$0.5 & +0.1   \\
$\gamma^{(p)}_4$ & +1.3  & +3.4   & +3.4  & +2.9   \\
\tableline
$\gamma^{(n)}_1$ & +4.4   & +6.1   & +5.9   & +6.1    \\
$\gamma^{(n)}_2$ &$-$0.3  &$-$0.8  &$-$1.0  & $-$0.9  \\
$\gamma^{(n)}_3$ & +1.1   &$-$0.6  &$-$0.6  & $-$0.6  \\
$\gamma^{(n)}_4$ & +1.3   & +3.4   & +3.5   & +3.6  \\
\end{tabular}
\end{center}
\end{table}
\vspace{0.5cm}

%
%
\vspace{0.5cm}
\begin{table}
\caption{Combinations of spin polarizabilities that do not depend on the
amplitude $A_2(\nu,t)$  (all results are in units of  
$10^{-4}\,$ fm$^4$).}
\begin{center}
\begin{tabular}{c|cccd}
$\gamma$'s & HBChPT  & DR (HDT) & $\text{SAID}_1$ & $\text{SAID}_2$ \\
\tableline
$\gamma^{(p)}$      & +2.0    &$-$0.6  &$-$1.2   &$-$1.3  \\
$\gamma^{(p)}_{13}$ & +6.6    & +3.8   & +3.3    & +3.7  \\
$\gamma^{(p)}_{14}$ & +1.8    &$-$1.7  &$-$2.4   &$-$2.3  \\
\tableline
$\gamma^{(n)}$      & +2.0    &$+$0.0 &$-$0.2  &$-$0.3    \\
$\gamma^{(n)}_{13}$ & +6.6    & +4.9  & +4.7   & +4.9    \\
$\gamma^{(n)}_{14}$ & +1.8    &$-$0.7 &$-$1.2  &$-$1.1   \\
\end{tabular}
\end{center}
\end{table}

\begin{figure}
\centerline{
\epsfxsize=15.0cm
\epsfbox{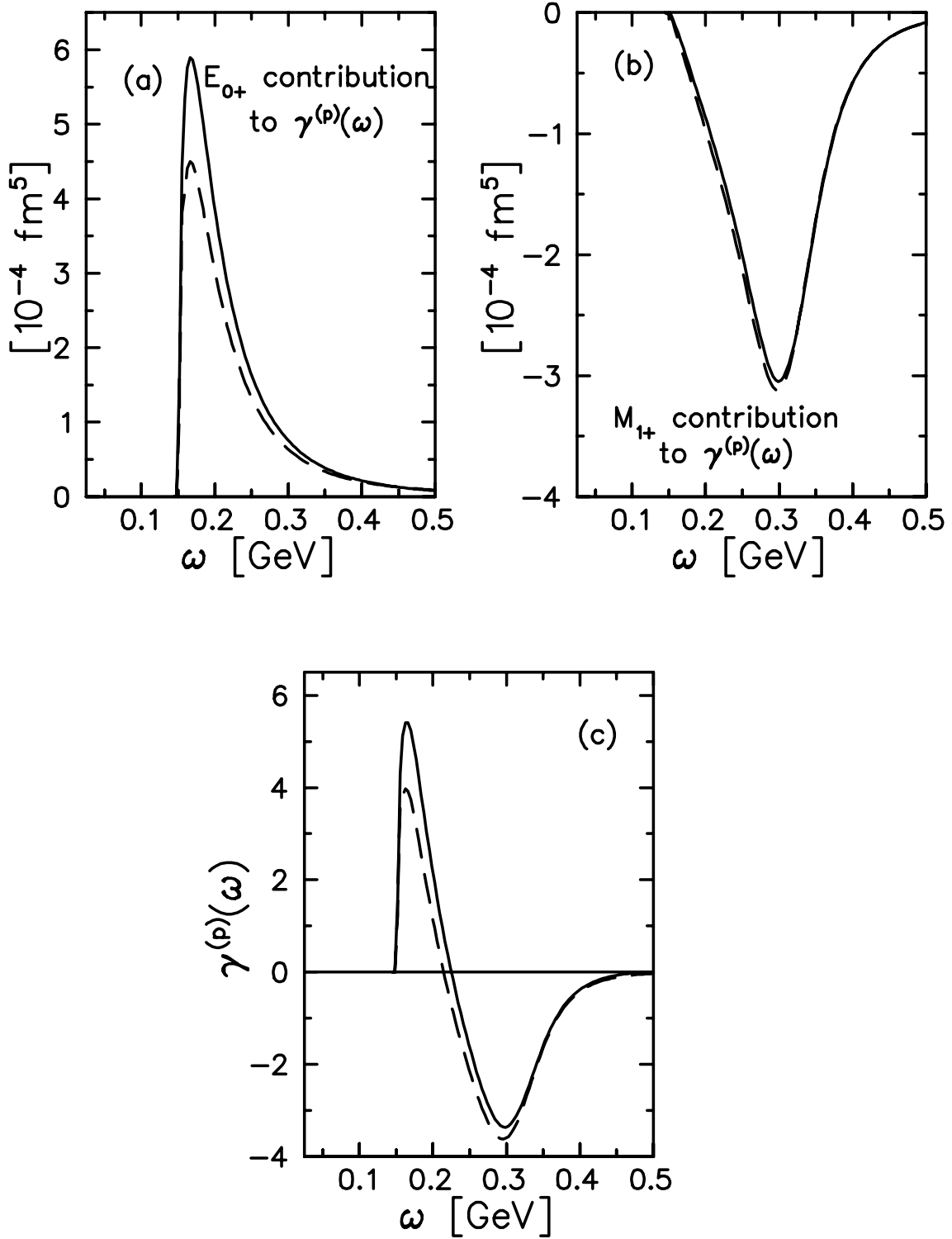}
}
\vspace{0.75cm}
\caption{Contribution of the multipoles $E_{0+}$ (a) and $M_{1+}$ (b) 
to the integrand for $\gamma^{(p)}$ (c) from the HDT (solid line) and the SAID 
(dashed line) analyses. }
\end{figure}
\end{document}